\documentclass[10pt, conference, compsocconf]{IEEEtran}
\ifCLASSINFOpdf
\else
\fi

\usepackage{times}
\usepackage{helvet}
\usepackage{courier}
\usepackage{subfigure}
\usepackage{amsfonts}
\usepackage{pifont}
\usepackage{amsmath}
\usepackage{graphicx}

\begin{document}

\newcommand{\rightarrowp}{\mathrel{\overset{+}{\longrightarrow}}}
\newcommand{\rightarrowm}{\mathrel{\overset{-}{\longrightarrow}}}
\newcommand{\leftarrowp}{\mathrel{\overset{+}{\longleftarrow}}}
\newcommand{\leftarrowm}{\mathrel{\overset{-}{\longleftarrow}}}

%
\title{A Model of Consistent Node Types in Signed Directed Social Networks}


\author{\IEEEauthorblockN{Dongjin Song}
\IEEEauthorblockA{Department of Electrical and Computer Engineering\\
University of California, San Diego\\
La Jolla, USA, 92093-0409\\
Email: \textsf{dosong@ucsd.edu}} \and \IEEEauthorblockN{David A.
Meyer}
\IEEEauthorblockA{Department of Mathematics\\
University of California, San Diego\\
La Jolla, USA, 92093-0112\\
Email: \textsf{dmeyer@math.ucsd.edu}} }


%


\maketitle

\begin{abstract}
Signed directed social networks, in which the relationships between
users can be either positive (indicating relations such as trust) or
negative (indicating relations such as distrust), are increasingly
common.  Thus the interplay between positive and negative
relationships in such networks has become an important research
topic.  Most recent investigations focus upon edge sign inference
using structural balance theory or social status theory.  Neither of
these two theories, however, can explain an observed edge sign well
when the two nodes connected by this edge do not share a common
neighbor (e.g., common friend).  In this paper we develop a novel
approach to handle this situation by applying a new model for node
types. Initially, we analyze the local node structure in a fully
observed signed directed network, inferring underlying node types.
The sign of an edge between two nodes must be consistent with their
types; this explains edge signs well even when there are no common
neighbors. We show, moreover, that our approach can be extended to
incorporate directed triads, when they exist, just as in models
based upon structural balance or social status theory.  We compute
Bayesian node types within empirical studies based upon partially
observed Wikipedia, Slashdot, and Epinions networks in which the
largest network (Epinions) has 119K nodes and 841K edges. Our
approach yields better performance than state-of-the-art approaches
for these three signed directed networks.
\end{abstract}

\begin{IEEEkeywords}
signed directed social networks; node types; Bayesian node features;
edge sign prediction.

\end{IEEEkeywords}

%
\IEEEpeerreviewmaketitle

\section{Introduction}
With the rapid emergence of social networking websites, e.g.,
Facebook, Twitter, LinkedIn, Epinions, etc., a considerable amount
of attention has been devoted to investigating the underlying social
mechanisms in order to enhance users' experiences
\cite{Liben-Nowell}\cite{Jiang}\cite{Leskovec2}\cite{Kadushin}.
Traditional social network analysis concerns itself primarily with
unsigned social networks such as Facebook or Myspace which can be
modeled as graphs, with nodes representing entities, and positively
weighted edges representing the existence of relationships between
pairs of entities. Recently, signed directed social networks, in
which the relationships between users can be either positive
(indicating relations such as trust) or negative (indicating
relations such as distrust), are increasingly common. For instance,
in Epinions \cite{Guha}, which is a product review website with an
active user community, users can indicate whether they trust or
distrust other users based upon their reviews; in Slashdot
\cite{Lampe}\cite{Brzozowski}, which is a technology-related news
website, users can tag each other as ``friend'' or ``foe'' based
upon their comments. Such a signed directed network can be modeled
as a graph expressed as an asymmetric adjacency matrix in which an
entry is $1$ (or $-1$) if the relationship is positive (or negative)
and 0 if the relationship is absent.

One of the fundamental problems in signed social network analysis is
edge sign inference \cite{Guha}\cite{Leskovec1}, i.e., inferring the
unknown trust or distrust relationship given the existence of a
particular edge. To address this issue, many approaches have been
developed based upon two main social-psychological theories, i.e.,
structural balance theory \cite{Heider}\cite{Cartwright} and social
status theory \cite{Leskovec}. Structural balance theory is more
well-known and it states that people in signed networks tend to
follow the rules that ``the friend of my friend is my friend'',
``the enemy of my friend is my enemy'', etc. Social status theory,
which is implicit in Guha et al.~\cite{Guha}, further exploited by
Leskovec et al.~\cite{Leskovec}, and based upon a foundation in
social psychology \cite{Kadushin}, considers a positive directed
edge to indicate that the initiator of the edge views the recipient
as having higher status and a negative directed edge to indicate
that the recipient is viewed as having lower status. The relative
levels of status determine the allowed sign-direction pairs for an
edge assuming that this edge exists.

Although both structural balance theory and social status theory
have proved useful for explaining the signs of edges in signed
networks, neither is suitable for explaining an observed edge when
the two nodes connected by this edge share no common neighbor (e.g.,
common friend), and in fact, structural balance theory simply does
not apply to this situation. Since many real world social networking
graphs tend to be very sparse, this is the case for a large fraction
of their edges. To better explain the observed edge signs in
general, in this paper we develop a novel approach to address this
issue by applying a new model for node types.

To summarize the contributions of this paper:
\begin{itemize}
  \item We explore the underlying local node structures in fully observed signed
    directed networks, recognizing that there are 16 different types of node and each type of
    node constrains both its incoming node types and its
    outgoing node types, i.e., the signs of their edges must be
    consistent with their types.
  \item We show that node type features can be extended to
    incorporate structural balance theory or social status theory,
    to help make predictions for those edges whose endpoints have
    common neighbors.
  \item For the purpose of practical applications, we derive
  Bayesian node features (including Bayesian node type and Bayesian node
  properties) based upon partially observed signed directed networks. 
  \item We conduct empirical studies based upon three real world datasets and
    show that our proposed approach can outperform state-of-the-art algorithms.
\end{itemize}
\section{Related Work}
\begin{figure}[b]
\centering\vspace{-5mm}
\includegraphics[angle=270,,scale=0.33]{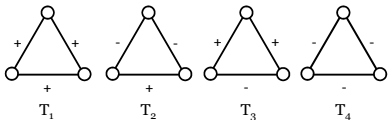}\vspace{-3mm}
\caption{Undirected signed triads. Structural balance theory states
that $\text{T}_1$ and $\text{T}_2$ are balanced, while $\text{T}_3$
and $\text{T}_4$ are unbalanced. Weak structural balance theory
states that $\text{T}_1$, $\text{T}_2$, and $\text{T}_4$ are
balanced, while only $\text{T}_3$ is
unbalanced.}\label{fig:fig2}\vspace{0mm}
\end{figure}

In the past few years, many approaches have been developed to
explore different aspects of signed networks, ranging from edge sign
prediction \cite{Guha}\cite{Leskovec1}\cite{Kai}\cite{Cho} to
community detection \cite{Kunegis}\cite{Anchuri}. Most of these
approaches are based upon structural balance theory or social status
theory.

\subsection{Structural balance theory}
The investigation of signed networks
\cite{Heider}\cite{Cartwright}\cite{Frank}\cite{Davis} can be traced
back to the 1920s. Heider~\cite{Heider} first formulated structural
balance theory within social psychology. After that, Cartwright and
Harary~\cite{Cartwright} formally provided the notion of structural
balance with undirected triads (as shown in Figure \ref{fig:fig2})
and proved its necessity and sufficiency by utilizing the
mathematical theory of graphs. Intuitively, their theory can be
explained as: ``the friend of my friend is my friend''
($\text{T}_1$), ``the enemy of my friend is my enemy''
($\text{T}_2$), ``the friend of my enemy is my enemy''
($\text{T}_2$), and ``the enemy of my enemy is my friend''
($\text{T}_2$). Conceptually, their theory claims that $\text{T}_1$
and $\text{T}_2$ are balanced while $\text{T}_3$ and $\text{T}_4$
are unbalanced. Davis~\cite{Davis} further generalized this theory
to weak structural balance theory by allowing all the edges of
triads to be negative, i.e., ``the enemy of my enemy is my enemy''
($\text{T}_4$ is also balanced). Note that these two balance
theories were initially intended for modeling undirected networks,
although they have been commonly applied to directed networks by
disregarding the direction of edges \cite{Leskovec}.

\subsection{Social status theory}

Guha et al.~\cite{Guha} first considered the edge sign prediction
problem by developing a trust propagation framework to predict the
trust (or distrust) between pairs of nodes. In their framework, they
calculate a combined matrix which is a linear combination of four
different one-step propagations, i.e., direct propagation,
co-citation, transpose trust, and trust coupling. Then the trust and
distrust propagations are achieved by calculating a linear
combination of powers of this combined matrix. A shortcoming of this
approach is that it cannot be explained by structural balance theory
\cite{Cartwright}\cite{Heider}.

\begin{figure}[t]
\centering\vspace{-2mm}
\includegraphics[angle=270,,scale=0.35]{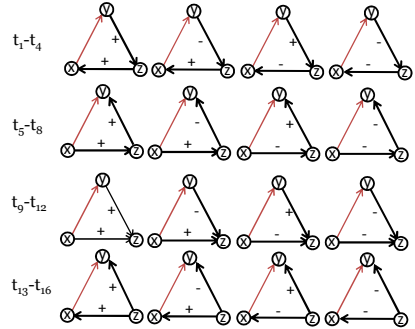}\vspace{0mm}
\caption{All contexts of ($x$, $y$; $z$). The red edge's sign is not
available; it can be determined based upon $x$'s and $y$'s
interactions with $z$. Take $t_1$ for example: since $y$ gives $z$ a
positive evaluation and $z$ gives $x$ a positive evaluation, $x$
tends to give $y$ a negative evaluation because $x$ has higher
status.}\label{fig:fig3}\vspace{0mm}
\end{figure}

Motivated by this trust propagation idea \cite{Guha} and informed by
social psychology \cite{Kadushin}, Leskovec et al.~\cite{Leskovec}
developed social status theory to explain signed directed networks.
In this theory, they assume that if there is a positive edge from
$x$ to $y$, it represents the fact that $x$ regards $y$ as having
higher status than himself (or herself), and if there is a negative
edge from $x$ to $y$, it represents the fact that $x$ regards $y$ as
having lower status than himself (or herself). Assuming everyone in
the system agrees on the same status ordering, we can infer signs
easily as long as the existence and direction of edges are
available. When prior status information for $x$ and $y$ is not
available, we can still perform sign inference using the context
provided by the rest of the network. For instance, in Figure
\ref{fig:fig3}, the sign of $x$ to $y$ can be inferred by referring
to the status of $z$, and is unambiguous in half the cases.

\subsection{Approaches to edge sign prediction}
Based upon structural balance theory and social status theory,
Leskovec et al.~\cite{Leskovec1} selected degree features and
directed triad features for edges in signed directed networks.
Specifically, for the edge from node $x$ to node $y$, they consider
seven degree features, i.e., $d_{\text{in}}^{+}(y)$ and
$d_{\text{in}}^{-}(y)$, the number of incoming positive and negative
edges to $y$, respectively; $d_{\text{out}}^{+}(x)$ and
$d_{\text{out}}^{-}(x)$, the number of outgoing positive and
negative edges from $x$, respectively; $C(x,y)$, the number of
common neighbors (i.e., embeddedness) of node $x$ and node $y$;
$d_{\text{out}}^{+}(x)+d_{\text{out}}^{-}(x)$ and
$d_{\text{in}}^{+}(y)+d_{\text{in}}^{-}(y)$, the total out-degree of
$x$ and the total in-degree of $y$, respectively. Since each of the
16 triad types in Figure \ref{fig:fig3} provides different evidence
for the sign of the edge from node $x$ to node $y$, directed triad
features of this edge are encoded in a 16-dimensional vector
counting the number of triads of each type in which this edge is
involved. After computing the degree or directed triad features for
the edge from $x$ to $y$, a logistic regression classifier is used
to combine the evidence from these individual features into an edge
sign prediction.

Subsequently, Chiang et al.~\cite{Kai} extended this approach by
considering longer cycles (e.g., quadrilaterals, pentagons) while
ignoring the directions of edges to reduce the computational
complexity. Hsieh et al.~\cite{Cho} formulated the sign inference
problem as a low rank matrix completion (approximation) problem
based upon weak balance theory. Note that this approach was
originally developed to explain a signed undirected network which is
associated with a symmetric adjacency matrix and is different from
our setting (signed directed networks) in this paper.

We remark that structural balance theory is also popular for
community detection in signed networks \cite{Kunegis}\cite{Anchuri}.

Although many approaches based upon structural balance theory or
social status theory have been developed to perform edge sign
prediction in signed networks, they cannot work well when few
topological features, i.e., undirected (or directed) triads and
long-range cycles, are available in the network. Since many real
world signed directed social networking graphs are very sparse, the
efficacy of methods based upon these theories is limited. A more
general approach for such networks is necessary.

\begin{table}[t]
\center
\begin{small}
\caption{Dataset statistics.} \centering{
\begin{tabular}{l||r|r|r}
\hline\hline
Datasets & Wikipedia  & Slashdot & Epinions \\
\hline\hline Nodes           & 7,118   & 82,144 & 119,217  \\
Edges           & 103,747   & 549,202 & 841,372  \\
$+$edges        & 78.78$\%$   & 77.4$\%$ & 85.0$\%$ \\
$-$edges        &  21.21$\%$  & 22.6$\%$ & 15.0$\%$   \\
\hline \hline
\end{tabular}
       \label{tab:table1}
} 
\end{small}\vspace{0mm}
\end{table}

\section{Datasets}

In this paper, we consider three well-known signed directed social
networks: Wikipedia \cite{Burke}, Slashdot
\cite{Kunegis2009}\cite{Lampe} and Epinions \cite{Guha}
\footnote{These datasets are available online at \hfill\break {\tt
http://snap.stanford.edu/data/}.}:

\begin{itemize}
  \item The Wikipedia data comprise a voting network for promoting
candidates to the role of admin. The voters, half coming from
existing admins and another half coming from ordinary Wikipedia
users, can indicate a positive (for supporting) or negative (for
opposing) vote with respect to the promotion of a candidate
\cite{Leskovec1}.
  \item Slashdot is a social website focusing on
technology related news. In Slashdot Zoo, users can tag each other
as friends (like) or foes (dislike) based upon comments on articles.
  \item Epinions, which is a product review website, is a trust network in
which users can indicate whether they trust or distrust each other
based upon their reviews.
\end{itemize}

The detailed statistics of these datasets are provided in
Table~\ref{tab:table1}. Note that in all three datasets, the
majority of the edges is positive. Due to this imbalance, simply
predicting all edges to be positive would yield 78.78$\%$, 77.4$\%$,
and 85.0$\%$ accuracy across the three datasets. To show the
effectiveness of any approach, it should achieve substantially
better performance than this.

\begin{figure}[t]
  \centering
  \begin{subfigure}[b][Wikipedia]{
  \centering
   \includegraphics[width=0.13\textwidth, angle=0] {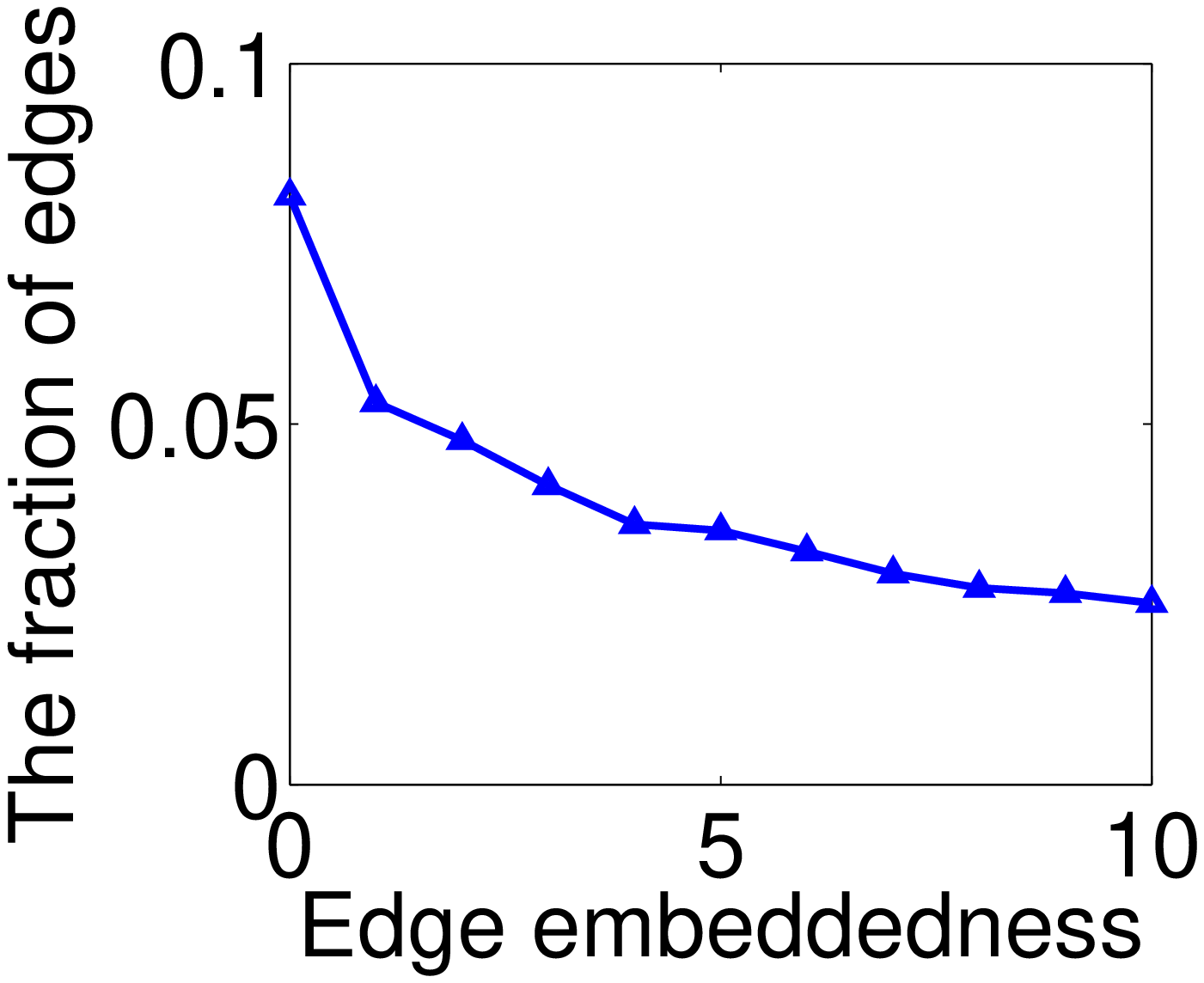}
   \label{fig:subfig4a}
 }
\end{subfigure}
\begin{subfigure}[b][Slashdot]{
  \centering
   \includegraphics[width=0.13\textwidth,angle=0] {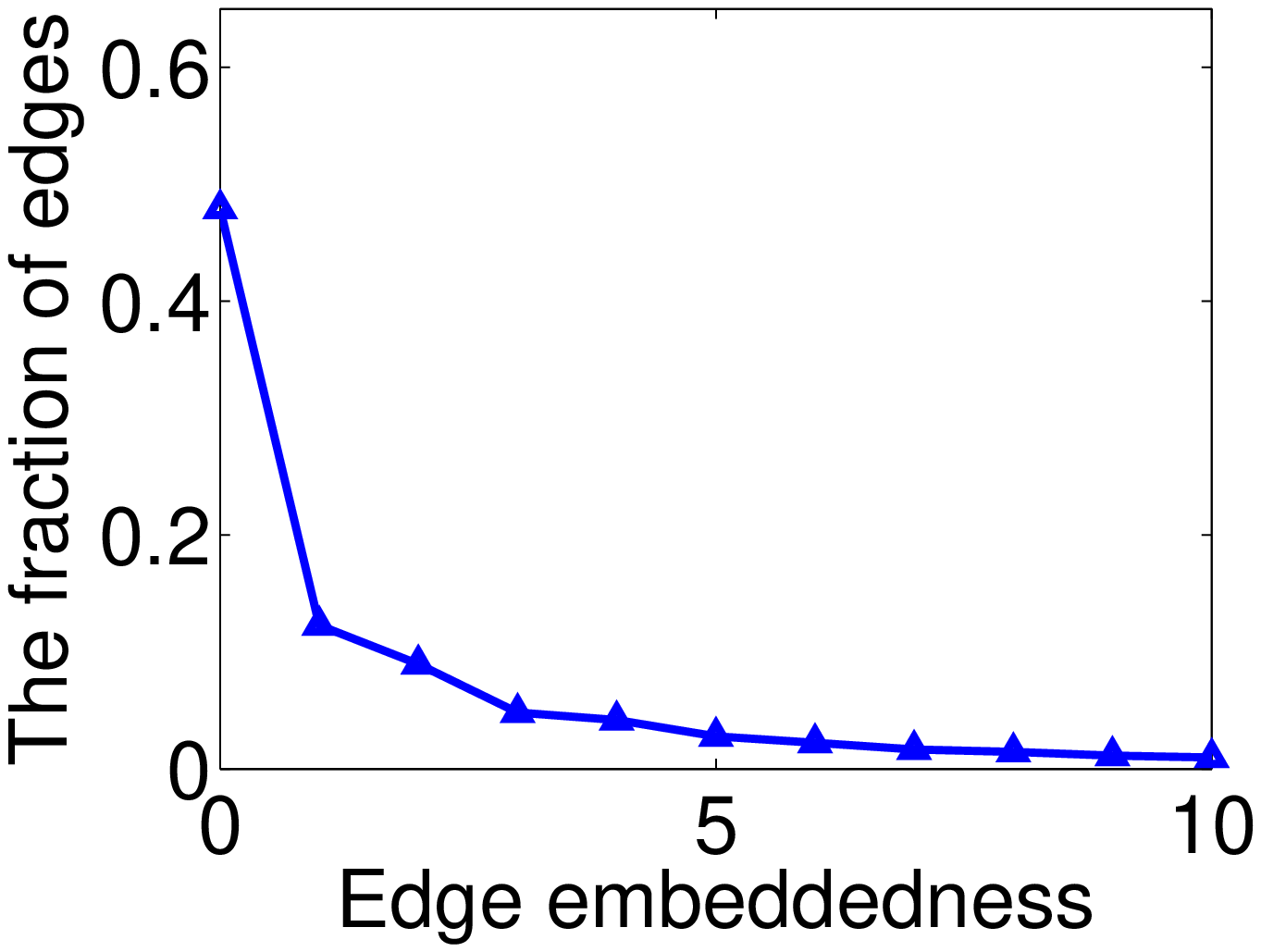}
   \label{fig:subfig4b}
 }
\end{subfigure}
\begin{subfigure}[b][Epinions]{
  \centering
   \includegraphics[width=0.13\textwidth,angle=0] {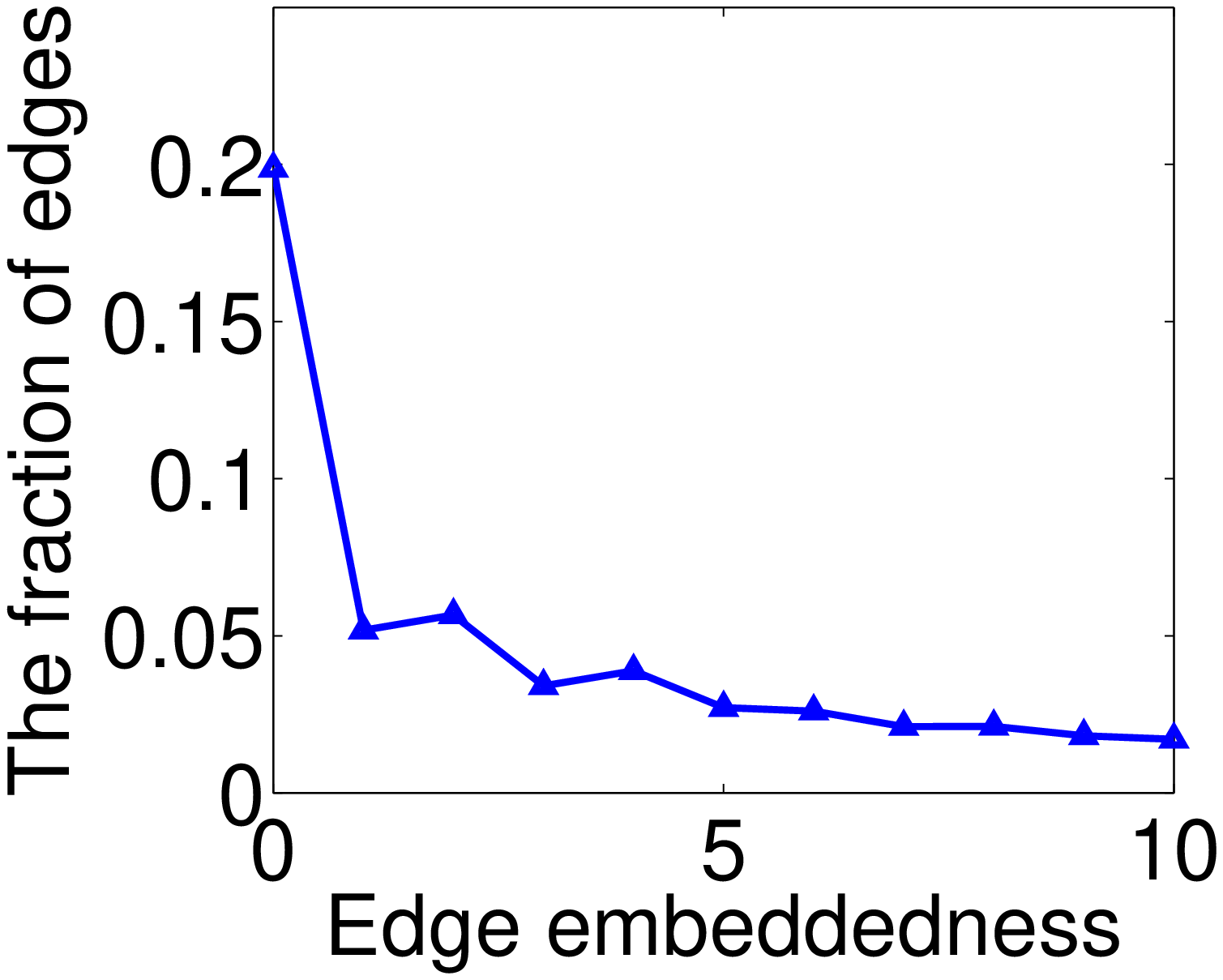}
   \label{fig:subfig4c}
 }
\end{subfigure}
    \caption{The fraction of edges vs. edge embeddedness for three
datasets.}\label{fig:4}\vspace{-1mm}
\end{figure}

Figure \ref{fig:4} shows the fraction of edges versus edge
embeddedness \cite{Granovertter} (the number of common neighbors of
the two nodes connected by the edge) for three datasets. We observe
that the edges with zero embeddedness comprise about $8.17\%$,
$47.90\%$, $19.88\%$ of the edges for Wikipedia, Slashdot, and
Epinions, respectively. Note that a large fraction of zero
embeddedness edges means that triad features \cite{Leskovec1} cannot
work well for edge sign prediction. This is because the entries of
triad feature vector will be zero and thus the triad features
provide no evidence for edge sign prediction.

\section{Node Types in Fully Observed Networks}

In this paper, a fully observed signed directed network refers to a
network in which there is no uncertainty about the existence of any
directed edge and its associated sign.

We consider a fully observed signed directed network as a graph
$G=(V,E,W)$, where $V$ is the vertex set of size $n$, $E$ is the
edge set of size $m$, and $W\in\mathbb{R}^{n \times n}$ is the
associated signed adjacency matrix. Because $G$ is a directed
network, $W$ is an asymmetric matrix and can be represented as:
\begin{align}
 W_{ij}=\left\{
  \begin{array}{ll}
  \vspace{0mm}
    1, & \hbox{if $i$ trusts $j$} \\
    -1, & \hbox{if $i$ distrusts $j$}\\
    0, & \hbox{otherwise}
\vspace{0mm}
  \end{array}
\right.
\end{align}
Note that $ W_{ij}=0$ represents no directed edge from node $i$ to
node $j$.

In this section, we first investigate local node structures within
fully observed signed directed networks, recognize a set of node
types, show that these node types can be used to explain real world
signed directed social networks, and show how to encode a specific
node in such networks.  Next, we explore how these node types
interact with one another and how these interactions can explain the
edge signs.  Finally, we show our approach can be extended to
incorporate structural balance theory or social status theory.


\begin{figure}[t]
\centering\vspace{0mm}
\includegraphics[angle=270,,scale=0.40]{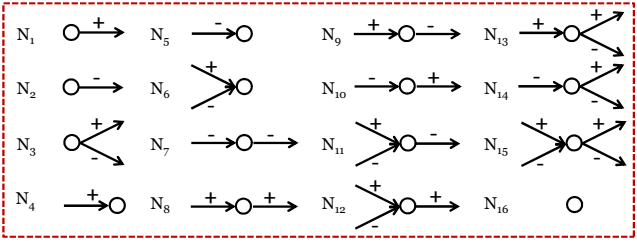}\vspace{3mm}
\caption{16 node types in signed directed networks. For $\text{N}_1$
(node type 1), all the outgoing edges are positive. For
$\text{N}_5$, all the incoming edges are negative. For
$\text{N}_{15}$, both the incoming edges and outgoing edges are
mixtures of positive edges and negative edges.}\label{fig:fig5}
\end{figure}

\subsection{Node types}

\begin{figure}[t]
\vspace{0mm}\centering
\includegraphics[angle=270,,scale=0.35]{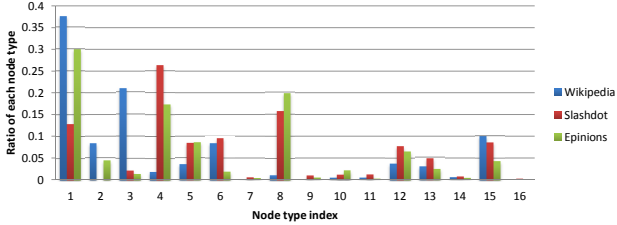}\vspace{3mm} \caption{The
fractions of different node types for the three
datasets.}\label{fig:fig6}
\end{figure}

In our study, we focus on analyzing signed directed networks because
they are more general and common than signed undirected networks in
real world applications. For instance, each of the three datasets we
consider in this paper is a signed directed network. Generally,
signed directed networks are sparse graphs in which nodes may be
categorized into 4 groups based upon whether they have incoming
edges and outgoing edges, i.e., nodes with neither incoming nor
outgoing edges (e.g., $\text{N}_{16}$ in Figure \ref{fig:fig5}),
nodes with only incoming edges (e.g., $\text{N}_{4}$,
$\text{N}_{5}$, and $\text{N}_{6}$ in Figure \ref{fig:fig5}), nodes
with only outgoing edges (e.g., $\text{N}_{1}$, $\text{N}_{2}$, and
$\text{N}_{3}$ in Figure \ref{fig:fig5}), and nodes having both
incoming and outgoing edges (e.g., $\text{N}_{9}$, $\text{N}_{14}$,
$\text{N}_{15}$, etc.). Moreover, both the incoming edges and the
outgoing edges of a given node can be categorized into 3 classes,
one class with only positive edges, another class with only negative
edges, and the third class with a mixture of positive and negative
edges. Combining these two principles, the nodes in signed directed
networks can be categorized into 16 types, shown in Figure
\ref{fig:fig5}. Note that the edges in Figure \ref{fig:fig5} only
indicate the types of the incoming or outgoing, i.e., they do not
represent the actual nonzero number of incoming (outgoing) positive
(negative) edges. The fractions of each node type for the three real
world datasets are shown in Figure \ref{fig:fig6}.

\subsubsection{The representation of node features}

To represent each node effectively, in addition to its node type
$\text{N}$, we should consider its associated node properties, i.e.,
the relative level of the number of positive (negative) incoming
edges $d_{\rm in}(+)$ $\big(d_{\rm in}(-)\big)$, and the relative
level of the number of positive (negative) outgoing edges $d_{\rm
out}(+)$ ($d_{\rm out}(-)$).

First, we can use a 16-dimensional binary vector,
$[\textbf{1}(\text{N}=\text{N}_{1})$,$\textbf{1}(\text{N}=\text{N}_{2})$,\ldots,
$\textbf{1}(\text{N}=\text{N}_{16})]$, to indicate the node type.
The indicator $\textbf{1}(\text{N}=\text{N}_{i})$ is 1 if $\text{N}$
is the same as $\text{N}_{i}$. Next, we can use a vector
$[\text{P}_{\rm in}(+), \text{P}_{\rm in}(-),$ $\text{P}_{\rm
out}(+), \text{P}_{\rm out}(-)]$ to denotes the ratio of positive
(negative) incoming edges and that of positive (negative) outgoing
edges.

\begin{figure}[t]
\vspace{0mm}\centering
\includegraphics[angle=270,,scale=0.40]{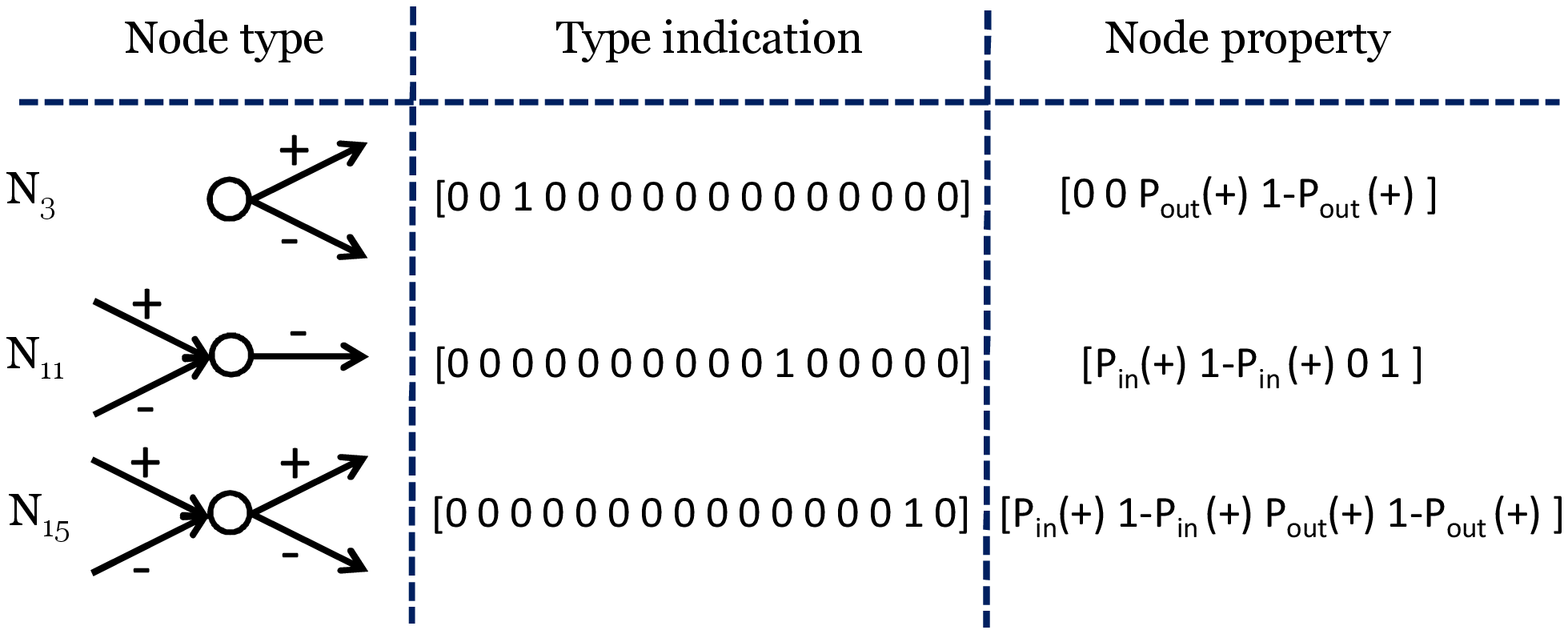}\vspace{3mm}
\caption{Two parts
of the features for node types $\text{N}_3$, $\text{N}_{11}$, and
$\text{N}_{15}$. Notice that the first (last) two numbers in the
second part need not sum to $1$.}\label{fig:fig7}
\end{figure}

Intuitively, $\text{P}_{\rm in}(+)$, $\text{P}_{\rm in}(-)$,
$\text{P}_{\rm out}(+)$, and $\text{P}_{\rm out}(-)$ represent the
locally propagating properties of a node (node property) and they
can be calculated with
\begin{align}
\text{P}_{\rm in}(+)=\frac{d_{\rm in}(+)}{d_{\rm in}(+)+d_{\rm
in}(-)+\varepsilon}
\end{align}
\begin{align}
\text{P}_{\rm in}(-)=\frac{d_{\rm in}(-)}{d_{\rm in}(+)+d_{\rm
in}(-)+\varepsilon}
\end{align}
\begin{align}
\text{P}_{\rm out}(+)=\frac{d_{\rm out}(+)}{d_{\rm out}(+)+d_{\rm
out}(-)+\varepsilon}
\end{align}
\begin{align}
\text{P}_{\rm out}(-)=\frac{d_{\rm out}(-)}{d_{\rm out}(+)+d_{\rm
out}(-)+\varepsilon},
\end{align}
where we set $\varepsilon=10^{-10}$ to avoid zero denominators.
Figure \ref{fig:fig7} shows examples of these two parts of features
for node type $\text{N}_3$, $\text{N}_{11}$, and $\text{N}_{15}$.
Notice that $\text{P}_{\rm in}(+)+\text{P}_{\rm in}(-)=1$ if there
is any input edge, but this sum is zero if there are none; so these
features are not redundant. Also note that node property is
essentially different from degree features because degree features
aim to model a particular edge by considering the initiator's
outgoing edges, recipient's incoming edges, and their common
neighbors.

Note that although the node properties, i.e., $[\text{P}_{\rm
in}(+), \text{P}_{\rm in}(-),$$\text{P}_{\rm out}(+), \text{P}_{\rm
out}(-)]$ implicitly indicate the node type information, it is still
useful to consider type indication, i.e.,
$[\textbf{1}(\text{N}=\text{N}_{1})$,$\textbf{1}(\text{N}=\text{N}_{2})$,\ldots,
$\textbf{1}(\text{N}=\text{N}_{16})]$. Since the latter is not a
linear combination of the former, it can provide non-redundant
information in the logistic regression classifier we will describe
in the Section 6.

\subsection{The interaction of node types}

We have shown there are 16 possible node types in any signed
directed network. Hence, theoretically there are $16^{2}$
combinations of node types. Given a node of a certain type, however,
it usually can only connect to (or be reached from) nodes in a
subset of these types due to the compatibility of both directions
and signs. In other words, there exists a logic to determine whether
two nodes can be reached or not and whether the sign should be
positive or negative. For instance, given a node of type
$\text{N}_5$, it can only be connected from a node of type
$\text{N}_2, \text{N}_3, \text{N}_7, \text{N}_9, \text{N}_{11},
\text{N}_{13}, \text{N}_{14},$ or $\text{N}_{15}$ and the edge sign
can only be negative. Similarly, given a node of type $\text{N}_9$,
it can only be connected from a node of type $\text{N}_1,
\text{N}_3, \text{N}_8, \text{N}_{10}, \text{N}_{12}, \text{N}_{13},
\text{N}_{14},$ or $\text{N}_{15}$ and connected to a node of type
$\text{N}_5, \text{N}_6, \text{N}_7, \text{N}_{10}, \text{N}_{11},
\text{N}_{12}, \text{N}_{14}$, or $\text{N}_{15}$. Moreover, the
edge sign is determined as positive and negative, respectively.

Given an edge from $x$ to $y$, based upon the combinations of node
type $x$ and node type $y$, this edge can be categorized into three
classes. $``+"$ denotes the edge sign is determined to be positive,
$``-"$ denotes the edge sign is determined to be negative, and
$``?"$ denotes the edge sign that cannot be determined by the
interaction of current two node types, i.e., the edge sign can be
either positive or negative. In our three datasets, there are
$29945$, $250487$ and $500309$ determined edges (i.e., $``+"$ and
$``-"$) for Wikipedia, Slashdot and Epinions, respectively, each a
large fraction of the total edges.

Although node types have shown their effectiveness for explaining
the edge signs in fully observed signed social networks, there
exists a fraction of the total edges for which signs cannot be
explained simply by node types. In this case, we can incorporate
structural balance or social status theory with node types to
address this issue.


\subsection{Incorporating structural balance or social status theory}

As we described the node types and the interactions of these node
types in the previous subsections, we did not need to consider
whether there is any common neighbor for a pair of nodes. We should,
however, be aware that when common neighbors exist for a pair of
nodes, structural balance or social status theory may help to
explain the sign of an edge between them.

\begin{figure}[t]
\vspace{0mm}\centering
\includegraphics[angle=270,,scale=0.40]{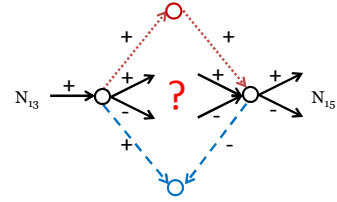}\vspace{0mm}
\caption{Examples for incorporation of structural balance theory or
social status theory into the node types interaction.
}\label{fig:fig8}\vspace{-4mm}
\end{figure}

In Figure \ref{fig:fig8}, for instance, since $\text{N}_{13}$ has
both positive and negative outgoing edges and $\text{N}_{15}$ has
both positive and negative incoming edges, the sign of the edge
between them cannot be determined by the interaction of these two
types. Since $\text{N}_{13}$ and $\text{N}_{15}$ have two common
neighbors, however, the sign of the edge between them may be
explained by either structural balance theory or social status
theory. Within structural balance theory, we can disregard the
directions of these two triads. From the red (dotted) triad, we can
infer the sign of the edge between $\text{N}_{13}$ and
$\text{N}_{15}$ to be positive based upon the rule that ``my
friend's friend is my friend''. From the blue (dashed) triad, on the
contrary, we can infer the sign of this edge to be negative based
upon the rule that ``my friend's enemy is my enemy''. Within social
status theory, since both the red (dotted) triad and the blue
(dashed) triad indicate that $\text{N}_{13}$ has higher status than
$\text{N}_{15}$, they consistently imply that the sign of the edge
from $\text{N}_{13}$ to $\text{N}_{15}$ is negative.

To incorporate these directed triads as features, we use the same
approach as Leskovec et al.~\cite{Leskovec1}\cite{Leskovec}. Given
an edge from $x$ to $y$, and a common neighbor $z$ of $x$ and $y$,
the edge between $x$ and $z$ can have four possible configurations,
i.e., $x \rightarrowp z$, $x \rightarrowm z$, $x \leftarrowp z$, and
$x \leftarrowm z$. Similarly, there are four possible signed edges
between $z$ and $y$. Hence we can obtain 16 types of triads each of
which may provide different evidence about the sign of the edge from
$x$ to $y$.

\section{Bayesian Node Features in Partially \\Observed Networks}

In real world applications, signed directed networks are often
partially observed, i.e., several edges' signs are unknown or
hidden. For example, in the Wikipedia dataset, we probably know that
someone has voted on a candidate, but we may not know this voter's
opinion. In this case, we would like to infer this voter's opinion
by learning some patterns based upon observed edges in the network.
However, when these unobserved edges take different signs, both the
node types and node properties may change. In this case, simple node
features (including node types and node properties) may not be
capable of capturing the range of possible unobserved signs and thus
will be not reliable.

To address this issue, we extend node features to Bayesian node
features by considering prior knowledge about unobserved signs in
partially observed signed directed networks.

Similarly to a fully observed signed directed network, a partially
observed signed directed network can also be represented as a graph
$G=(V,E,W)$, where $V$ is the vertex set of size $n$, $E$ is the
edge set of size $m$, and $W\in\mathbb{R}^{n \times n}$ is the
associated signed adjacency matrix. Since $G$ is a directed network,
$W$ is an asymmetric matrix and can be represented as:
\begin{align}
 W_{ij}=\left\{
  \begin{array}{ll}
  \vspace{0mm}
    1, & \hbox{if $i$ trusts $j$} \\
    -1, & \hbox{if $i$ distrusts $j$}\\
    $?$, & \hbox{an edge from $i$ to $j$ exists, but sign is unknown}\\
    0, & \hbox{otherwise}
\vspace{0mm}
  \end{array}
\right.
\end{align}
$ W_{ij}=0$ represents no directed edge from node $i$ to node $j$.

In this section, we first introduce Bayesian node type, show how to
calculate it based upon both observed (training) edges and
unobserved (test) edges, and present two ways to encode the
interaction of Bayesian node types. Next, we show how to calculate
and represent Bayesian node properties.

\subsection{Bayesian node type}

\begin{figure}[t]
\vspace{2mm}\centering
\includegraphics[angle=270,,scale=0.30]{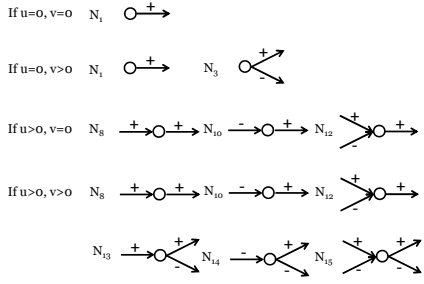}\vspace{-1mm} \caption{The possible
node of type 1 with different $u$ and
$v$.}\label{fig:fig9}\vspace{-6mm}
\end{figure}

Given any node in a partially observed network, assuming $u$ denotes
the number of unobserved incoming edges and $v$ denotes the number
of unobserved outgoing edges, let $P_{u}(+)$ (or $P_{u}(-)$)
represent the prior probability of incoming edges being positive
(negative) and $P_{v}(+)$ (or $P_{v}(-)$) represent the prior
probability of outgoing edges being positive (negative). From these
probabilities, we can calculate its probability distribution over
node types.

Take a node of type $\text{N}_{1}$ for example (as shown in Figure
\ref{fig:fig9}), if $u=0$ and $v=0$, its node type does not change;
if $u=0$ and $v>0$, it has $(P_{v}(+))^{v}$ probability to be
$\text{N}_{1}$ and $1-(P_{v}(+))^{v}$ probability to be
$\text{N}_{3}$;
 if $u>0$ and $v=0$, it has $(P_{u}(+))^{u}$ probability to be
$\text{N}_{8}$, $(P_{u}(-))^{u}$ probability to be $\text{N}_{10}$,
and $1-(P_{u}(+))^{u}-(P_{u}(-))^{u}$ probability to be
$\text{N}_{12}$; if $u>0$ and $v>0$, it has
$(P_{u}(+))^{u}(P_{v}(+))^{v}$ probability to be $\text{N}_{8}$,
$(P_{u}(-))^{u}(P_{v}(+))^{v}$ probability to be $\text{N}_{10}$,
$(1-(P_{u}(+))^{u}-(P_{u}(-))^{u})(P_{v}(+))^{v}$ probability to be
$\text{N}_{12}$, $(P_{u}(+))^{u}(1-(P_{v}(+)^{v}))$ probability to
be $\text{N}_{13}$, $(P_{u}(-))^{u}(1-(P_{v}(+)^{v}))$ probability
to be $\text{N}_{14}$, and $(1-(P_{u}(+))^{u}-(P_{u}(-))^{u})$
$(1-(P_{v}(+)^{v}))$ probability to be $\text{N}_{15}$. This
calculation determines a 16 dimensional vector which encodes the
distribution of possible node types. Note that similar vectors can
be calculated for other types of nodes. We do not specify the
calculation for each node type due to the the space limit.

To initialize $P_{u}(+)$ and $P_{v}(+)$, we can also use the
Bayesian node properties $P_{\text{in}}(+)$ and $P_{\text{out}}(+)$,
i.e., $P_{u}(+)=P_{\text{in}}(+)$ and $P_{v}(+)=P_{\text{out}}(+)$,
to claim that each unobserved edge obeys Bayesian node properties
(i.e., local priors).

Given an observed (training) edge connecting node $x$ and node $y$,
we can obtain two vectors $V_{x}\in\mathbb{R}^{16}$ and
$V_{y}\in\mathbb{R}^{16}$ by calculating their Bayesian node types.
To encode the interaction of Bayesian node types, we can (1) simply
concatenate these two vectors to form a 32 dimensional vector; or
(2) calculate the Kronecker product of these two vectors, i.e.,
$V_{x}\bigotimes V_{y}$, and form a 256 dimensional vector. We
should be aware that the vector formed by the Kronecker product
encodes the probability distribution of different node type
interactions.

\begin{figure}[t]
\vspace{-5mm}\centering
\includegraphics[angle=0,,scale=0.35]{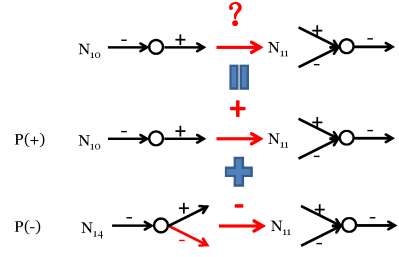} \vspace{-7mm}\caption{Example of
calculating the Bayesian node type interaction when two nodes are
connected with an unobserved edge. Note that the Bayesian node types
of $\text{N}_{10}$, $\text{N}_{11}$,and $\text{N}_{14}$ should be
calculated similar with $\text{N}_{1}$ in Figure \ref{fig:fig9}.
}\label{fig:fig10}\vspace{-6mm}
\end{figure}

Given an unobserved (test) edge connecting node $x$ and node $y$, we
should consider both possible signs as shown in Figure
\ref{fig:fig10}. Specifically, we first decompose this kind of
interaction into two separate cases, i.e., the edge sign being
positive and negative. Next, we calculate the Bayesian node types,
represent their interactions (either concatenation vector or
Kronecker product vector) of both cases. Finally, we calculate the
linear combination of these two cases with respect to the prior
probability of the signs (i.e., $P(+)$ and $P(-)$) over observed
edges.

\subsection{Bayesian node properties}

Given a node in a partially observed network, assuming $u$ denotes
the number of unobserved incoming edges and $v$ denotes the number
of unobserved outgoing edges, by assigning to these $u+v$ edges
different signs, the node properties also change.

To capture the range of possible signs of unobserved edges, we
should consider Bayesian node properties, i.e., incorporating prior
information, namely the expected number of incoming positive
(negative) and outgoing positive (negative) edges, with the number
of positive (negative) incoming edges $d_{\rm in}(+)$ $\big(d_{\rm
in}(-)\big)$ or outgoing edges $d_{\rm out}(+)$ $\big(d_{\rm
out}(-)\big)$.

Specifically, the Bayesian node properties are represented as
following:
\begin{align}
\text{P}_{\rm in}(+)=\frac{d_{\rm in}(+)+P(+)u}{d_{\rm in}(+)+d_{\rm
in}(-)+u+\varepsilon}
\end{align}
\begin{align}
\text{P}_{\rm in}(-)=\frac{d_{\rm in}(-)+P(-)u}{d_{\rm in}(+)+d_{\rm
in}(-)+u+\varepsilon}
\end{align}
\begin{align}
\text{P}_{\rm out}(+)=\frac{d_{\rm out}(+)+P(+)v}{d_{\rm
out}(+)+d_{\rm out}(-)+v+\varepsilon}
\end{align}
\begin{align}
\text{P}_{\rm out}(-)=\frac{d_{\rm out}(-)+P(-)v}{d_{\rm
out}(+)+d_{\rm out}(-)+v+\varepsilon},
\end{align}
where $P(+)$ is the prior probability of positive edges and $P(-)$
is the prior probability of negative edges.

To encode the interaction of Bayesian node properties, we simply
concatenate two of the Bayesian node properties vectors to form an 8
dimensional vector.

As in the previous section, when common neighbors exist for a pair
of nodes, structural balance or social status theory may be
incorporated with Bayesian node features to help explain the sign of
an edge between them.

\section{Supervised Learning of the \\Proposed Features}

Given a fully observed signed directed network, the node type
interactions, extended to triads with structural balance or social
status theory, are useful to explain the edge signs. Partially
observed signed directed networks, however, are too complicated to
fully conform to the rules of simple node type interactions. Also,
as illustrated in Figure \ref{fig:fig8}, if there are multiple
common neighbors for $\text{N}_{13}$ and $\text{N}_{15}$, structural
balance theory (or social status theory) may conflict with itself.
To address this issue, we can utilize a logistic regression to
combine the evidence from the interaction of Bayesian node features
and triad features.

We now consider the features collected for the logistic regression.
The features we utilize can be divided into three classes. One class
comes from Bayesian node type interaction (32 or 256 dimensional
vector); another class is based upon Bayesian node properties
interaction (8 dimensional vector); the last class is triads (16
dimensional vector).

Given a partially observed signed directed social network, we first
use a logistic regression to fit the features of observed edges
(training data) and then utilize the learned coefficients to
linearly combine the evidence from each individual feature of
unobserved edges (test data) so as to predict the sign. The logistic
regression can be written in the following form
\begin{align}
P(\ell=1|\textbf{f})=\frac{1}{1+\exp[-(\textbf{w}^{\rm{T}}\textbf{f}+w_{0})]}
\end{align}
where $\ell\in\{0,1\}$ is the label, $1$ represents positive edge
while $0$ represents negative edge. $\textbf{f}\in\mathbb{R}^{d}$ is
the feature vector, and $[\textbf{w}; w_{0}]\in\mathbb{R}^{d+1}$ are
the coefficients we estimate from the features of observed edges
(training data).

\begin{figure*}[t]
  \centering
  \begin{subfigure}[b][Wikipedia]{
  \centering
   \includegraphics[width=0.27\textwidth, angle=0] {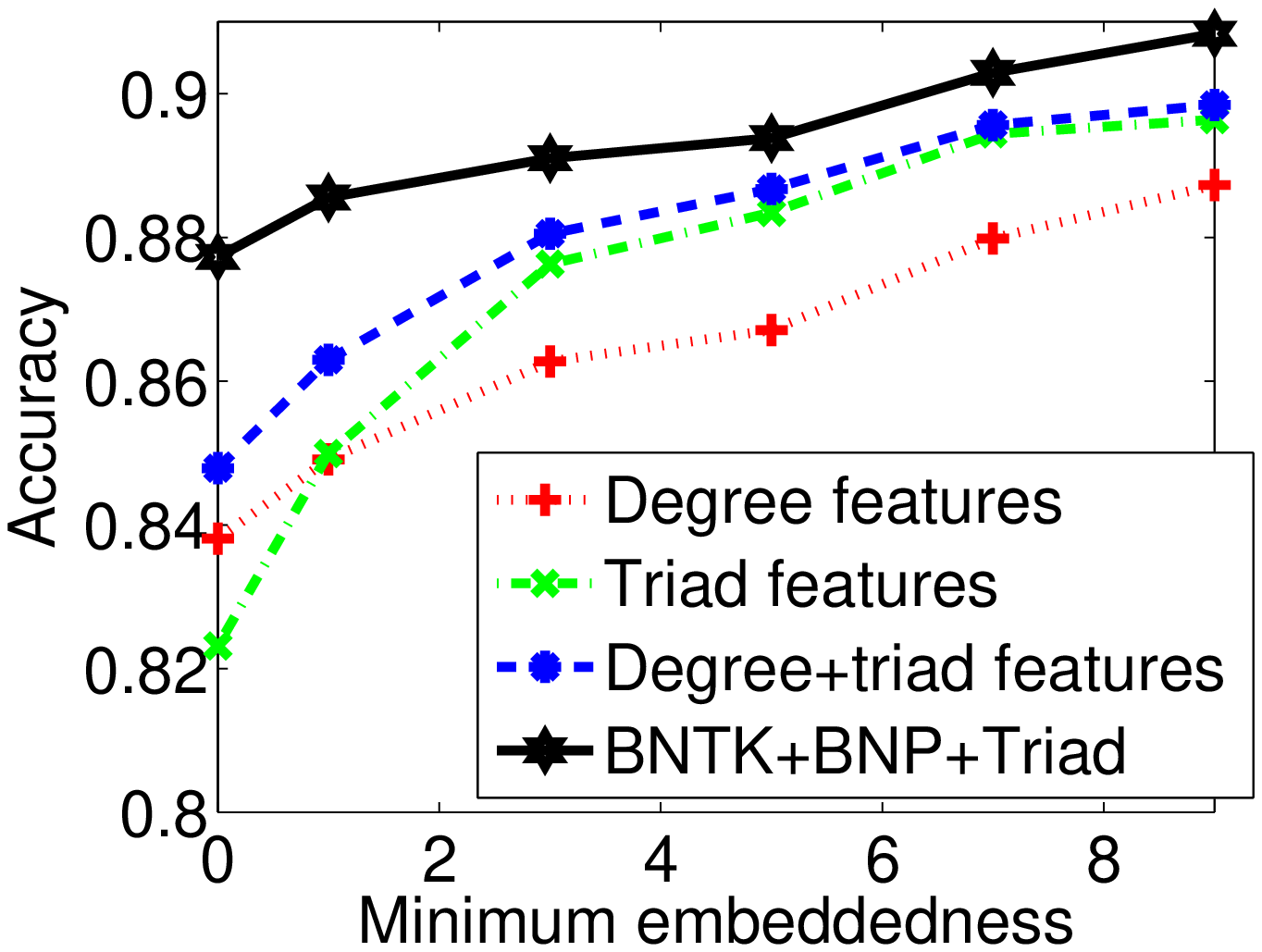}
   \label{fig:subfig10a}
 }
\end{subfigure}
\begin{subfigure}[b][Slashdot]{
  \centering
   \includegraphics[width=0.27\textwidth,angle=0] {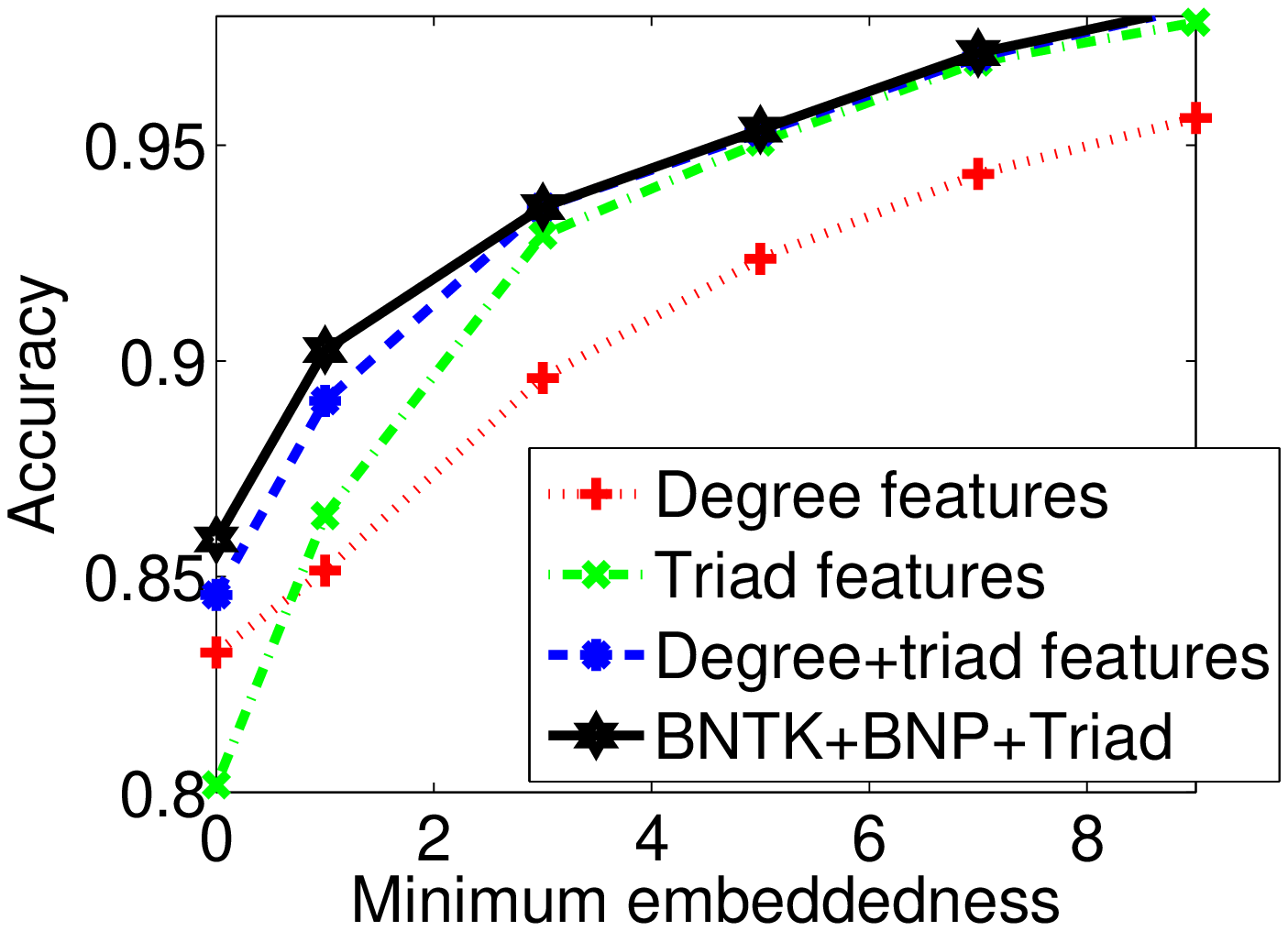}
   \label{fig:subfig10b}
 }
\end{subfigure}
\begin{subfigure}[b][Epinions]{
  \centering
   \includegraphics[width=0.27\textwidth,angle=0] {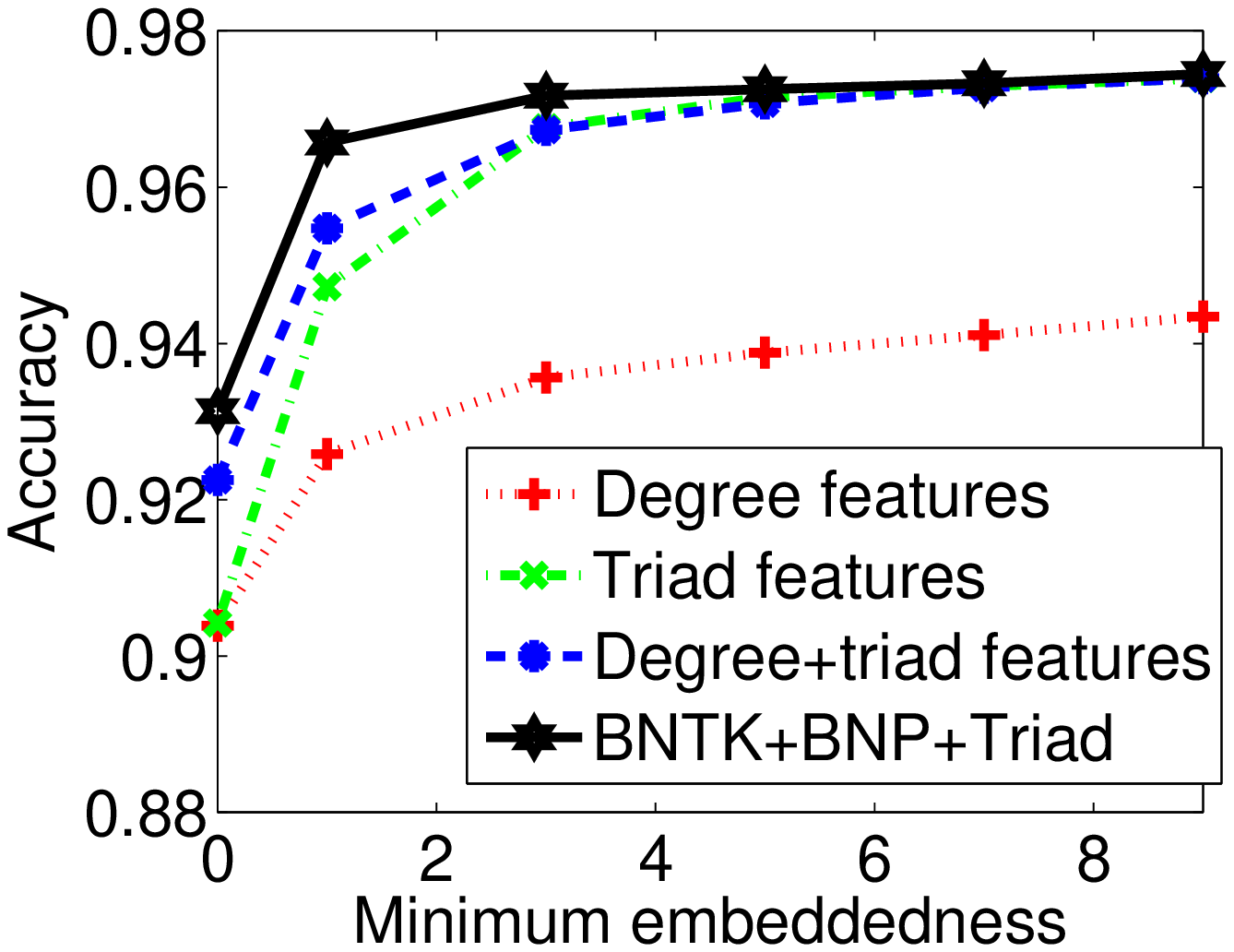}
   \label{fig:subfig10c}
 }
\end{subfigure}
    \caption{The edge sign prediction accuracies of different approaches vs. minimum embeddedness.}\label{fig:fig11}\vspace{-5mm}
\end{figure*}

\section{Experiment}

In this section, we conduct empirical studies based upon  Wikipedia
\cite{Burke}, Slashdot \cite{Kunegis2009}\cite{Lampe} and Epinions
\cite{Guha}. We first construct three fully observed asymmetric
adjacency matrices (as in Eq.(1)) based upon these three datasets.
Next, for each adjacency matrix, we randomly remove 10$\%$ of edges'
signs and form a partially observed network (as in Eq. (6)).
Subsequently, we calculate Bayesian node features (including
Bayesian node types and Bayesian node properties) and triad features
for both observed (training) and unobserved (test) edges. Then, we
estimate the parameters of logistic regression based upon the
features of observed edges and make predictions based upon the
features of unobserved edges. In our experiment, we repeat this
procedure 5 times and report the average prediction accuracy and
standard deviation for each approach. The baseline approaches are
implemented with identical parameter settings as in the original
works for fair comparisons.

\begin{table}[t]
\centering \centering
 \caption{Step by step justification results on Wikipedia, Slashdot, and Epinions (accuracy: $\%$  and standard deviation (std): $\%$).} \label{tab:table4}
 \vspace{1mm}
\begin{tabular}{l||c|c|c}
  \hline\hline
     &Wikipedia & Slashdot & Epinions\\
\hline
BNTC           & 83.55($\pm$0.56) & 80.32($\pm$0.15) & 89.65($\pm$0.04)\\
BNTK           & 83.84($\pm$0.12) & 80.89($\pm$0.08) & 88.34($\pm$0.29) \\
BNTC+BNP       & 87.03($\pm$0.15) & 84.48($\pm$0.06) & 92.96($\pm$0.03) \\
BNTK+BNP       & 86.98($\pm$0.19) & 84.90($\pm$0.09) & 92.46($\pm$0.03) \\
BNTC+BNP+Triad & 87.28($\pm$0.26) & 85.24($\pm$0.11) & \textbf{93.61($\pm$0.02)}\\
BNTK+BNP+Triad & \textbf{87.37($\pm$0.22)} & \textbf{85.65($\pm$0.11)} & 93.13($\pm$0.04) \\
  \hline \hline
\end{tabular}\vspace{-5mm}
\end{table}

\subsection{Step by step justification}

We examine the effectiveness of the proposed features by testing
each component step by step. We use BNTC to represent encoding the
interaction of Bayesian node types with concatenation, use BNTK to
represent encoding the interaction of Bayesian node types with the
Kronecker product, use BNP to represent the interaction of Bayesian
node properties, and use Triad to denote triad features
\cite{Leskovec1}\cite{Leskovec}.

Table~\ref{tab:table4} shows the results of step by step
justification for edge sign prediction on three datasets. We observe
that the interaction of Bayesian node types (BNTC and BNTK)
generally outperforms simply predicting all edges to be positive.
This demonstrates that the interaction of Bayesian node types is
useful to explain the edge signs in partially observed social
networks. We also observe that encoding Bayesian node types with the
Kronecker product achieves better performance than concatenation on
Wikipedia and Slashdot, while concatenation perform slightly better
on Epinions.

By concatenating BNTC and BNTK with Bayesian node properties (BNP)
features, we observe that BNTC+BNP and BNTK+BNP consistently
outperforms BNTC and BNTK. This is because Bayesian node properties
(BNP) provide more specific information about the incoming positive
(negative) and outgoing positive (negative) edges of nodes.

Finally, we show that, by concatenating BNTC+BNP and BNTK+BNP with
triad features to form BNTC+BNP+Triad and BNTK+BNP+Triad, the
performances are consistently slightly improved. This is because
triad features are useful to explain the edge signs when common
neighbors are available.

The step by step justification not only examines the effectiveness
of each component of the proposed Bayesian node features, but also
shows that the Bayesian node features can incorporate structural
balance or social status theory in the form of triad features.

\subsection{Edge sign prediction}

In this subsection, we compare Bayesian node features (including
Bayesian node types and Bayesian node properties) plus triad
features with state-of-the-art approaches, i.e., degree features
\cite{Leskovec1}, triad features \cite{Leskovec1}, degree$+$triad
features \cite{Leskovec1}, longer cycles features \cite{Kai}, and
low rank modeling \cite{Cho}. Note that in our experiment we extract
longer cycles features based upon the partially observed asymmetric
adjacency matrix and report the best performance over order 3, 4,
and 5 for comparison. Also notice that low rank modeling \cite{Cho}
can only theoretically analyze the undirected signed networks; in
our experiment, we adapt it and apply it to partially observed
signed directed networks.
\begin{table}[t]
\center \caption{Edge sign prediction accuracies (accuracy: $\%$
standard deviation (std): $\%$).}\label{tab:table7} \vspace{-2mm}
\scriptsize
\begin{tabular}{l||c|c|c}
\hline\hline
Datasets & Wikipedia  & Slashdot & Epinions \\
\hline Degree features \cite{Leskovec1}                 & 83.58($\pm$0.60)   & 83.76($\pm$0.13) & 90.39($\pm$0.25) \\
\hline Triad features \cite{Leskovec}                   & 82.46($\pm$0.52)   & 80.42($\pm$0.21) & 90.42($\pm$0.13) \\
\hline Degree+triad features \cite{Leskovec1,Leskovec}  & 84.87($\pm$0.08)   & 84.91($\pm$0.02) & 92.25($\pm$0.15) \\
\hline Longer cycles features \cite{Kai}                & 84.04($\pm$0.39)   & 83.83($\pm$0.34) & 90.64($\pm$0.28)  \\
\hline Low rank modeling \cite{Cho}                     & 84.93($\pm$0.54)   & 84.57($\pm$0.46) & 92.48($\pm$0.32)  \\
\hline \hline BNTC+BNP+Triad                            & 87.28($\pm$0.26)   & 85.24($\pm$0.11) & \textbf{93.61($\pm$0.02)}\\
\hline BNTK+BNP+Triad                                   & \textbf{87.37($\pm$0.22)} & \textbf{85.65($\pm$0.11)} & 93.13($\pm$0.04) \\
\hline\hline
\end{tabular}\vspace{-5mm}
\end{table}

In Table \ref{tab:table7}, we compare BNTC+BNP+Triad and
BNTK+BNP+Triad with the other five state-of-the-art approaches. We
observe that BNTC+BNP+Triad and BNTK+BNP+Triad consistently
outperform the other five algorithms. Note that these two variants
achieve best accuracies of $87.37(\pm0.22)\%$, $85.65(\pm0.11)\%$,
and $93.61(\pm0.02)\%$ over Wikipedia, Slashdot and Epinions,
respectively. This is because these two variants not only can
explain the edge signs well when common neighbors are not available
but also can effectively explain the edge signs when common
neighbors exist.

In Figure \ref{fig:fig11}, we compare BNTK+BNP+Triad with degree
features, triad features, and degree$+$triad features at different
levels of embeddedness (the number of common neighbors). In general,
we observe that when the minimum embeddedness increases, the
performance of BNTK+BNP+Triad increases. This is because structural
balance theory and social status theory are incorporated into
BNTK+BNP+Triad in the form of triads (Triad) and are effective in
explaining edge signs when common neighbors exist. Moreover, we
notice that BNTK+BNP+Triad generally outperforms other methods with
different levels of embeddedness. This is because BNTK+BNP+Triad
leverages the power of node type interactions as well as the power
of structural balance or social status theory in the form of triad
features (Triad).

\subsection{Cross-dataset evaluation}
We conduct cross-dataset evaluation with degree features, triad
features, degree$+$triad features, and Bayesian node features plus
triad features in the form of BNTK$+$BNP$+$Triad on these three
datasets. The aim is to examine the generalization capability of
each approach. In particular, given each type of features, we train
them on one dataset (e.g., Wikipedia) and evaluate the edge sign
prediction performance on another (e.g., Slashdot). For each pair of
datasets, the test is conducted 5 times based upon the random
selected test sets. We report the average accuracies of different
approaches in Table \ref{tab:table8}.

We observe that Bayesian node features plus triad features in the
form of BNTK$+$BNP$+$Triad can achieve the best performance on each
pair of the cross-dataset evaluation. This illustrates that Bayesian
node features plus triad features not only are useful on
intra-dataset evaluation but also have good generalization
capability. This is extremely helpful for edge sign prediction in
signed networks with few training examples.

\section{Conclusions}

In this paper, we explored the underlying local node structures in
signed networks, recognizing that there are 16 different types of
node and each type of node constrains both its incoming node types
and its outgoing node types, i.e., the sign of an edge between two
nodes must be consistent with their types. This is a highly
structured alternative to the ordered scalar node types postulated
by social status theory. We demonstrated that the interaction
between these more complicated node types can explain edge signs
well. We also showed that our approach can be extended to
incorporate triad features whose inclusion is motivated by
structural balance theory or social status theory. We derived
Bayesian node features (including Bayesian node type and Bayesian
node properties) based upon partially observed signed directed
network. Empirical studies based upon three large scale datasets,
i.e., Wikipedia, Slashdot, and Epinions showed that the proposed
Bayesian node features plus triad features outperform
state-of-the-art algorithms on edge sign prediction. Moreover, we
showed that Bayesian node features plus triad features are more
effective than baseline approaches for cross-dataset edge sign
predictions.

In the future, it will be interesting to study the link
recommendation problem based upon Bayesian node features as well as
other explicit topological features of signed social networks.

\section*{Acknowledgment}
This work was partially supported by the Minerva Research Initiative
under ARO grants W911NF-09-1-0081 and W911NF-12-1-0389.
\begin{table}[t]
 \caption{Cross-dataset evaluation results. Training is conducted on the column datasets and testing is conducted on the row datasets. } \label{tab:table8} \vspace{-3mm}
\centering \subtable[Degree feature (accuracy: \%)]{ \scriptsize
\centering
\begin{tabular}{|c|ccc|}\hline
$\text{}$&$\text{Wikipedia}$&$\text{Slashdot}$&$\text{Epinions}$\\\hline
$\text{Wikipedia}$&83.58&83.87&92.66\\
$\text{Slashdot}$&80.34&83.76&90.79\\
$\text{Epinions}$&79.59&81.69&90.39\\\hline
\end{tabular}
\label{tab:chapter4:1a} } \subtable[Triad feature (accuracy: \%)]{
\scriptsize \centering
\begin{tabular}{|c|ccc|}\hline
$\text{}$&$\text{Wikipedia}$&$\text{Slashdot}$&$\text{Epinions}$\\\hline
$\text{Wikipedia}$&82.46&79.04&89.95\\
$\text{Slashdot}$&83.18&80.42&91.05\\
$\text{Epinions}$&81.66&79.19&90.42\\\hline
\end{tabular}
\label{tab:chapter4:1b} } \subtable[Degree+Triad feature (accuracy:
\%)]{ \scriptsize \centering
\begin{tabular}{|c|ccc|}\hline
$\text{}$&$\text{Wikipedia}$&$\text{Slashdot}$&$\text{Epinions}$\\\hline
$\text{Wikipedia}$&84.87&84.66&93.29\\
$\text{Slashdot}$&82.93&84.91&92.90\\
$\text{Epinions}$&81.96&83.07&92.25\\\hline
\end{tabular}
\label{tab:chapter4:1c} } \subtable[BNTK+BNP+Triad (accuracy: \%)]{
\scriptsize \centering
\begin{tabular}{|c|ccc|}\hline
$\text{}$&$\text{Wikipedia}$&$\text{Slashdot}$&$\text{Epinions}$\\\hline
$\text{Wikipedia}$&\textbf{87.37}&\textbf{85.21}&\textbf{93.53}\\
$\text{Slashdot}$&\textbf{87.31}&\textbf{85.65}&\textbf{93.23}\\
$\text{Epinions}$&\textbf{87.02}&\textbf{83.37}&\textbf{93.13}\\\hline
\end{tabular}
\label{tab:chapter4:1d} }\vspace{-5mm}
\end{table}


\begin{thebibliography}{1}
\bibitem{Anchuri}
P.~Anchuri and M.~Magdon-Ismail, \emph{Communities and balance in
signed networks: A spectral approach}, Proceedings of the 4th
IEEE/ACM ASONAM, August, 2012.

\bibitem{Brzozowski}
M.~J.~Brozzowski, T.~Hogg, and G.~Szabo, \emph{Friends and foes:
ideological social networking}, Proceedings of the ACM SIGCHI,
pp.817-820, Florence, Italy, 2008.

\bibitem{Burke}
M.~Burke and R.~Kraul, \emph{Mopping up: Modeling wikipedia
promotion decisions}, Proceedings of the ACM Conference on Computer
Supported Cooperative Work, pp.27-36, 2008.

\bibitem{Cartwright}
D.~Cartwright and F.~Harary, \emph{Structural balance: A
generalization of {H}eider's theory}, Psych. Rev., vol. 63, no. 5,
pp. 277-293, 1956.

\bibitem{Kai}
K.~Y.~Chiang, N.~Natarajan, A.~Tewari, and I.~S.~Dhillon,
\emph{Exploiting longer cycles for link prediction in signed
network}, Proceedings of the 20th ACM CIKM, pp.1,157-1,162, Glasgow,
Scotland, 2011.

\bibitem{Davis}
J.~A.~Davis, \emph{Clustering and structural balance in graphs},
Human Relations, vol. 20, no. 2, pp. 181-187, 1967.

\bibitem{Frank}
O.~Frank and F.~Harary, \emph{Balance in stochastic signed graphs},
Social Networks, vol. 2, pp. 155-163, 1979.

\bibitem{Guha}
R.~V.~Guha, R.~Kumar, P.~Raghavan, and A.~Tomkins, \emph{Propogation
of trust and distrust}, Proceedings of the 13th WWW, pp.403-412, New
York, NY, 2004.

\bibitem{Granovertter}
M.~Granovetter, \emph{Economic action and social structure: The
problem of embeddedness}, American Journal of Sociology, vol. 91,
no. 3, pp. 481-510, November, 1985.

\bibitem{Cho}
C.~J.~Hsieh, K.~Y.~Chiang, and I.~S.~Dhillon, \emph{Low rank
modeling of signed network}, Proceedings of the ACM SIGKDD, 2012.

\bibitem{Heider}
F.~Heider, \emph{Attitudes and cognitive organization}, Journal of
Psychology, vol. 21, pp. 107-112, 1946.


\bibitem{Jiang}
J.~Jiang, C.~Wilson, X.~Wang, P.~Huang, Y.~Dai, and B.~Y.~Zhao,
\emph{Understanding latent interactions in online social networks},
Proceedings of the ACM the Internet Measurement Conference,
November, 2010.

\bibitem{Kadushin}
C.~Kadushin, Understanding social networks: Theories, concepts, and
findings, \emph{Oxford University Press}, 2012.

\bibitem{Kunegis2009}
J.~Kunegis, A.~Lommatzsch, and C.~Bauckhage, \emph{The slashdot zoo:
Mining a social network with negative edges}, Proceedings of the
19th WWW, pp.541-550, 2009.

\bibitem{Kunegis}
J.~Kunegis, S.~Schmidt, A.~Lommatzsch, J.~Lerner, E.~W.~DeLuca, and
S.~Albayrak, \emph{Spectral analysis of signed graphs for
clustering, prediction and visualization}, Proceedings of the SIAM
SDM, pp.559-570, 2010.

\bibitem{Lampe}
C.~Lampe, E.~Johnston, and R.~Resnick, \emph{Follow the reader:
filtering comments on slashdot}, Proceedings of the ACM SIGCHI,
pp.1,253-1,262, San Jose, CA, 2007.

\bibitem{Liben-Nowell}
D.~Liben-Nowell and J.~Kleinberg, \emph{The link prediction for
social networks}, Journal of American Society for Information
Science and Technology, vol. 58, no. 7, pp. 1,019-1,031, 2007.

\bibitem{Leskovec1}
J.~Leskovec, D.~P.~Huttenlocher, and J.~M.~Kleinberg,
\emph{Predicting positive and negative links in online social
networks}, Proceedings of the 20th WWW, pp.641-650, 2010.

\bibitem{Leskovec}
J.~Leskovec, D.~P.~Huttenlocher, and J.~M.~Kleinberg, \emph{Signed
networks in social media}, Proceedings of the ACM SIGCHI,
pp.1,361-1,370, 2010.

\bibitem{Leskovec2}
J.~Leskovec, K.~J.~Lang, and M.~W.~Mahoney, \emph{Empirical
comparison of algorithms for network community detection},
Proceedings of the 20th WWW, 2010.

\bibitem{Yang11}
S.~H.~Yang, B.~Long, A.~Smola, N.~Sadagopan, Z.~Zheng, and H. Zha,
\emph{Like like alike - Joint friendship and interest propagation in
social networks}, Proceedings of the 21st WWW, pp.537-546, 2011.

\bibitem{Yang}
X.~Yang, H.~Steck, and Y.~Liu, \emph{Circle-based recommendation in
online social networks}, Proceedings of the ACM SIGKDD, 2012.



\end{thebibliography}
\end{document}